# Parsec structure and properties of the jet of 3C273. Results of Space VLBI data processing

A. Chuprikov* and I. Guirin†

*Astro Space Center of P. N. Lebedev Physical Institute of Russian Academy of Sciences, 117997, 84 / 32 Profsoyuznaya street, Moscow, Russia
†common address for author1 and author2

**Abstract.** We present result of processing of data of ground-space VLBI experiment titled *W068*. Particularly, one part of data of that observational session is successfully processed. These data were obtained on 2000 March 17 between 9:00 UT and 10:30 UT. 10 antennas of American interferometer *VLBA (Very Long Baseline Array)* and Japan satellite *VSOP (VLBI Space Observatory Programme* ([4])) were involved into this experiment. Moreover, 27 antennae of *VLA (Very Large Array)* were used as an additional ground antenna. Data were transferred from archive of the *NRAO (National Radio Astronomy Observatory, USA)* and processed with the software titled '*Astro Space Locator' (ASL for Windows)* (see [1]). The main result of this processing is the image of the quasar titled 3C273 with high resolution and high accuracy. Using this image, we make some conclusions about the radio structure of jet of this object. Our result is not in conflict with other results of processing of the Space VLBI data for 3C273 published earlier with many authors ([5], [6]). We could add some new aspects into that results. The reconstructed images of 3C273 for 6 centimeter wavelength range and values of some parameters of this source are presented.

## INTRODUCTION

Observations of many astrophysical objects with VLBI method could dramatically change our understanding of the nature of these objects. Most valuable are the VLBI observations on baselines whose length is about size of the Earth. The Japanese program VSOP uses the baselines with maximum baseline length. We present one result of processing of data of Space VLBI observation of the well-known quasar titled 3C273 (J1229+0203). For this quasar $z = 0.158$ (see [3]). Then, for $H_0 = 80 km \cdot s^{-1} \cdot Mpc^{-1}$ and $q_0 = 0.5$ value of distance is 580 Mpc and consequently *1 milliarcsecond in map corresponds to the linear size of approximately 2.9 pc*

## SOME PROPERTIES OF SPACE VLBI OBSERVATIONS

We take into consider a little part of the W068 observational session. 3C273 was observed on March 17, 2000 since 9:00 UT until 10:30 UT. As mentioned above, all the VLBA antennae as well as the VLA were used as a ground part of the interferometer. A 70-meter antenna in Tidbinbilla, Australia was used as a tracking station in order to recieve the signal from the VSOP satellite. Since only one polarization channel of satellite (L) was capable ([4]) we take into account only one type of polarization (LL).

**TABLE 1.** Parameters of orbit of two SVLBI missions satellites

|  | VSOP* | 'Radioastron'† |
|---|---|---|
| Semi-major axis | 17342.353 km | 189000.0 km |
| Period | 6.3 hours | 7 - 10 days |
| Eccentricity | 0.603 | 0.8 |
| Inclination | 31.45 deg | 51.6 deg |
| Longitude of ascending node | 185.1264 deg | - |
| Argument of perigee | 238.636 deg | - |
| Mean anomaly | 75.7474 deg** | - |

\* see http://www.vsop.isas.jp/NewOrbit.html
† see http://www.asc.rssi.ru/radioastron/description/orbit_eng.htm
\*\* for 18/07/2000 00:00 UT

Two 16 MHz bands totally covering the range between 4850 MHz and 4882 MHz were recorded. Each frequency band consists of 128 frequency channels. The primary data processing was made at the NRAO VLBA correlator in Socorro, New Mexico, USA.

Table 1 contains parameters of orbit of the VSOP satellite. For comparison, values of the same parameters for the 'Radioastron' mission orbit are also shown in the same table. Values of longitude of ascending node, and argument of perigee are still not established for the 'Radioastron' mission. Value of mean anomaly of the 'Radioastron' satellite for certain moment of time could be estimated when mission will be started. Thus, there are not these three values in table 1.

As we can see from this table, the maximum baseline length of the 'Radioastron' mission is at least 10 times more than the maximum baseline length of the VSOP. Thus, we hope to have the better resolution due to using of such orbit. Eccentricity of the 'Radioastron' orbit is also more than eccentricity of the VSOP orbit. Moreover, it is supposed that evolution of the 'Radioastron' orbit will be essential. Therefore, we hope to obtain the reliable coverage of (u, v)-plane even for observations near to the apogee. Then, the 'Radioastron' mission will have important advantages in comparison with the VSOP mission.

## METHOD AND RESULTS OF DATA PROCESSING

The software project Astro Space Locator (ASL for Windows) (see [1]) has been used to process all the data. The data processing consists of the following stages :
- Amplitude calibration of all the data using values from GC and TY tables and some additional information
- Single band fringe fitting (the primary phase calibration) of all the data. Estimation of the optimum value of solution interval
- Averaging of all the data over each frequency band
- Multi band fringe fitting of the atmosphere and ionosphere calibrator data. Estimation of gain values for every frequency and every time interval
- Application of gains, compensation of atmosphere and ionosphere delay for all the data

• Self-calibration. Final averaging, editing of the data, and *Imaging*

Amplitude calibration of the data was made with usage of two standard calibration tables (*'Gain Curve' (GC)* and *'System Temperature' (TY)*) that could be established, in principle, with the VLBA correlator during the primary data processing. Unfortunately, there were not calibration data for the VSOP antenna in both tables in initial FITS-file created with correlator. Moreover, such data for some VLBA antennae also were not available. Thus, it was necessary to find all the calibration information in corresponding NRAO archives. Additionally, we had to get the following values of calibration parameters for the VSOP satellite antenna (see http://www.vsop.isas.ac.jp/obs/HALCAcal.html for detailed information) :

1. Effective area value for C frequency range is 17 square meters. In other words, *Degrees per Flux Unit value (DPFU)* is equal to *0.0062 K/Jy*.
2. System temperature values of VSOP Channel (including the tracking station equipment) for C frequency range are the following :
   • 96 - 102 K for the lower band (4850 - 4866 MHz)
   • 102 - 106 K for the upper band (4866 - 4882 MHz)

Then, we have to write manually the text calibration file that contains all the values above. This file has been used for the ASL amplitude calibration procedure similar to Astronomical Image Processing System (AIPS) task 'ANCAL'. Result of application of this procedure to initial visibility function is shown in Figure 1. It is clear from this picture that the integration flux value for these frequencies is equal to approximately *43 Jy*. This value is consistent with data from *C Band VLA/VLBA Polarization Calibration Database* (see http://www.vla.nrao.edu/astro/calib/polar/2000/C_band_2000.shtml).

The main problem of any VLBI data processing is the reconstruction of the phase of visibility function. As mentioned above, two 16 MHz bands were recorded during the observations. First of all, it is necessary to perform the *fringe fitting* procedure for each band independently (*Single Band Fringe Fitting*, *SBFF*). As the baseline length is extremely big for SVLBI observations, result of this procedure could depend on the solution interval very strong. Our experience demonstrate that the best value of solution interval of the SBFF procedure for the *W068* observational session is about *30 seconds*. At least, we could guarantee the constancy of values of delay, fringe rate and initial phase for so short time interval. Then, after the SBFF procedure was successfully completed, all the data could be averaged on frequency as well as on time. The phase of visibility function versus the (u, v)-radius for 3C273 after application of the *SBFF* procedure to initial data with consequent data averaging in time and frequency is shown in Figure 2.

As both, visibility amplitude, and visibility phase have successfully been reconstructed (for first approximation), we could to perform the inverse Fourier transformation to obtain the *Dirty Map* of source. This Dirty Map is shown in Figure 3. It is necessary to take into account the following details of this image :
• a central region of image is quite symmetrical
• a maximum of intensity is slightly shifted from the centrum of the map

Thus, we could make the following conclusion : in order to reconstruct the phase of the visibility function for the Space VLBI data with required accuracy, it is not sufficient

322

to perform the ordinary amplitude calibration procedure with additional data of system temperatures and effective areas of all the antennae and to perform the ordinary fringe fitting procedure.

Then, in this case it is necessary to take into account the broad band effects. We could use two additional sources observed in these observations in order to calibrate the influence of the Earth atmosphere as well as the Earth ionosphere. These sources are *J1337-1257* and *J1256-0547 (3C279)*. They have the compact and simple structure in the C frequency range (see ([3])). Additionally, both sources are close to 3C273.

The main principle of our method of compensation of phase errors caused by media is that we use the Least Squares Method to solve some set of equations *simultaneously* for atmosphere and ionosphere. The *atmosphere* influence assumed to be independent on the frequency. Its dependence on time is described with polynomial function of second power. Another assumption is that the phase error caused by the *ionosphere only* can be expressed as :

$$\Phi_k(t,f) = Q_k(t)\frac{1}{f} \tag{1}$$

where $\Phi_k(t,f)$ is the visibility function phase error for k-th antenna
$Q_k(t)$ is a coefficient of proportionality of ionospere phase error to $\frac{1}{f}$
$f$ is a frequency [Hz]

It is important that $Q_k(t)$ is antenna based parameter and depends on the TEC (*Total Electronic Content*) value that is variable in time for every antenna independently each other. See ([2]) for some details.

Then, two calibrators, *J1337-1257* and *J1256-0547 (3C279)* were used jointly to estimate the $Q_k(t)$ function value as well as to estimate the atmosphere calibration polynomial function coefficients values for every antenna and every time interval. The data estimated are used to find the values of the *atmospheric/ionospheric gains*. The averaged visibility function must be divided by these values. Then, the data obtained will be free of the atmospheric/ionospheric errors.

The next stage of data processing is procedure of the *Self-Calibration*. Result of usage of this procedure is an accurate model of source structure as well as accurate values of gains for each antenna each frequency and each time interval. Usually, the *Corrected Visibility Function* is used for analysis of source structure. It is sufficient to divide the Initial Visibility Function by gains estimated with the Self-Calibration procedure in order to find all the values of this function. Sometimes, if quality of VLBI data is not so high, the *Total Model Map Visibility Function* could be used for such analysis. Such Total Model Map Visibility Function could be estimated easily if we know the source structure and the (u, v)-plane coverage.

The final resulting map of 3C273 central region is shown in Figure 4. We could note the following properties of this image :
• after final phase reconstruction, radio structure is not symmetrical. There is one-sided jet typical for quasar morphology
• radio structure is compact and simple. its sizes are 15 * 20 milliarcseconds. Hence, value of the linear sizes of this region is approximately 40 * 60 parsecs
• values of peak brightness, and r.m.s. level are 33.028 $Jy \cdot beam^{-1}$ , and 0.6 $Jy \cdot beam^{-1}$ correspondingly. Heance, the dynamic range value is about 55

**TABLE 2.** Parameters of components of radio structure of 3C273 (see Figure 4)

|   | Flux [Jy] | Relative RA [mas] | Relative DEC [mas] |
|---|-----------|-------------------|--------------------|
| 1 | 20.32     | 0.0               | 0.0                |
| 2 | 5.79      | -2.0              | -0.75              |
| 3 | 14.5      | -5.2              | -1.71              |
| 4 | 1.44      | -9.8              | -6.49              |
| 5 | 0.94      | -12.0             | -8.40              |
| 6 | 0.80      | -14.0             | -8.48              |

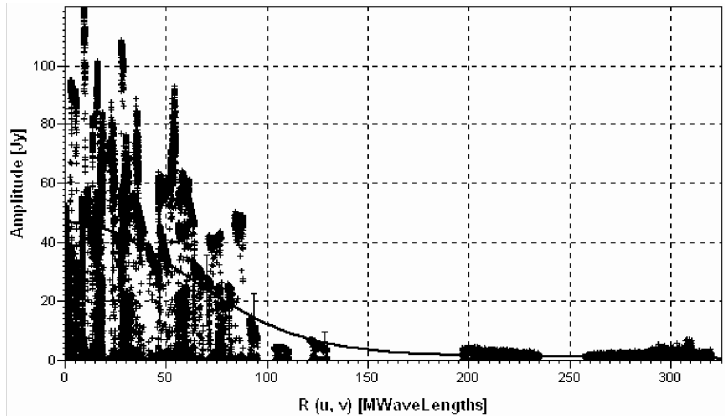

**FIGURE 1.** The visibility amplitude as a fuction of (u, v)-radius

- parameters of 6 brightest components of image are listed in table 2. The integration flux value was estimated for each of them with two dimensional integration of brightness distribution. Limits of integration correspond to the half of intensity of each component

It is useful to demonstrate the best resolution of Space VLBI in comparison with the ordinary VLBI. Two (u, v)-plane coverages are shown in Figure 5. It is clear from this picture, that maximum baseline projection length is about 5 times more for Space VLBI in comparison with the ordinary VLBI. On this reason, the resolution is essentially better for the Space VLBI case. It is shown in Figure 6.

# CONCLUSIONS

1. Part of data of archive Space VLBI experiment *W068* had been carried out in 2000 March is successfully processed with the software titled *'Astro Space Locator' (ASL for Windows)*
2. The advanced scheme of VLBI data processing is used. Particularly, we have used the method of simultaneous compensation of atmosphere and ionosphere delays with usage of data of two calibration sources close to the main sources titled 3C273

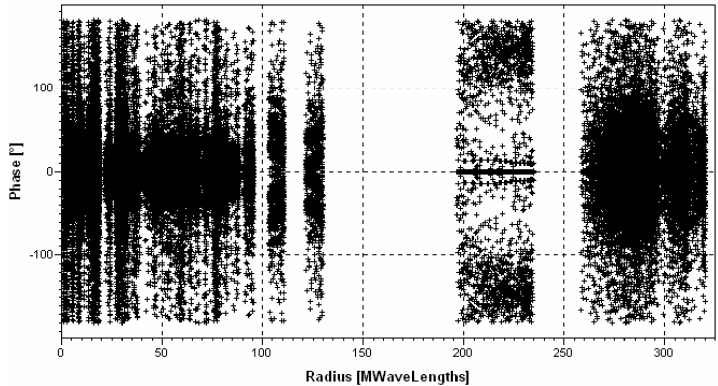

**FIGURE 2.** The visibility phase as a fuction of (u, v)-radius

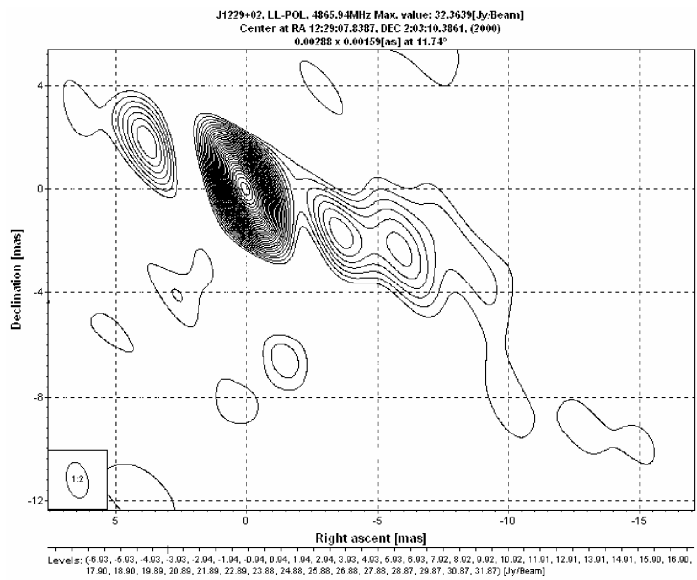

**FIGURE 3.** Dirty map. It was obtained immediately from amplitude (Figure 1) and phase (Figure 2) recostructed

3. It is demonstrated that the ordinary scheme of VLBI data processing is not relevant for Space VLBI data processing due to extremely big length of the ground-space baselines. The phase of visibility function must be reconstructed very carefully in this case
4. It is demonstrated that including of the orbital antenna could extend the (u, v)-plane essentially even if satellite is close to perigee of orbit. The advanced scheme of data

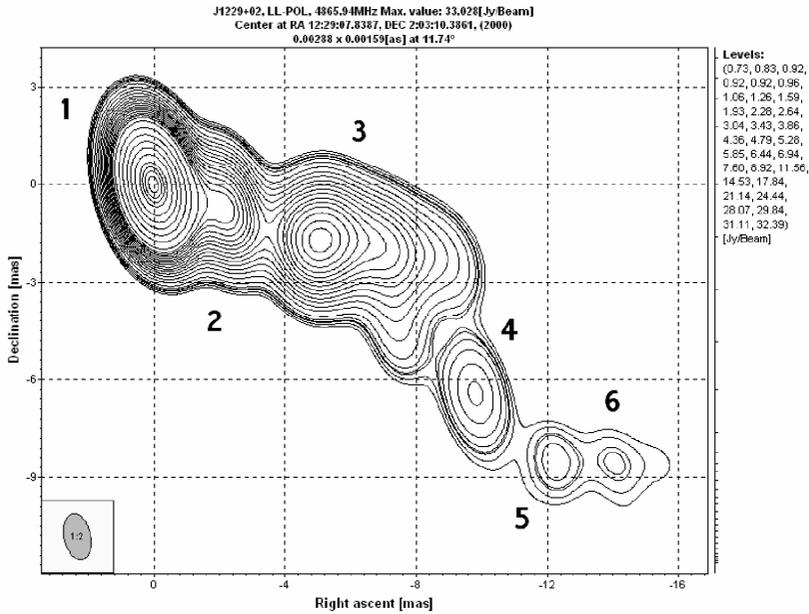

**FIGURE 4.** The final image reconstructed for 3C273

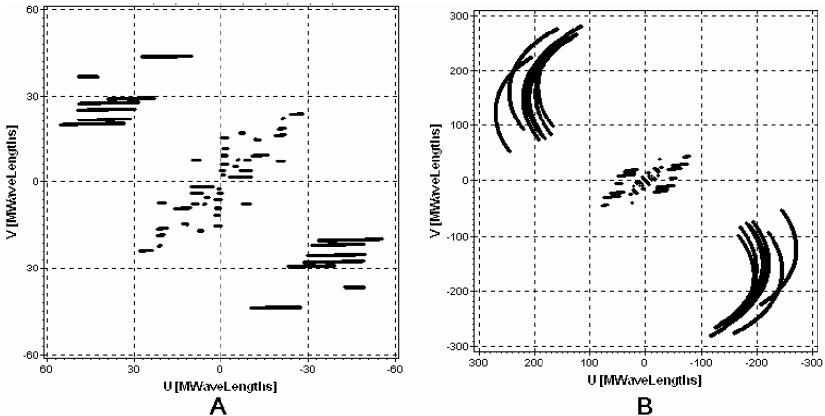

**FIGURE 5.** (u, v)-plane coverage for ground only (A) and space-ground interferometry (B)

processing allows to avoid the big decreasing of signal-to-noise ratio in this case
5. It is demonstrated that usage of the orbital antenna could reduce the beam sizes for about 2 - 3 times (for our case) and, consequently, improve the resolution essentially
6. As we can see from the 3C273 final map reconstructed, some components of central

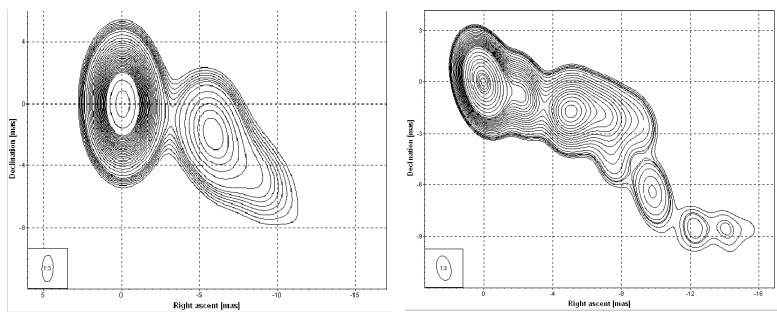

**FIGURE 6.** The image reconstructed for ground only (Left) and space-ground interferometry (Right)

region still remain non-resolved (particlarly, component number 3 in Figure 4). Then, we could confirm that 3C273 will be very interesting for other Space VLBI missions, particularly, for the 'Radioastron'

7. Procedures and technologies created and advanced during the *W068* data processing will be valid and relevant for the 'Radioastron' mission data processing

All the results presented in this paper are preliminary. It is necessary to process other parts of the *W068* observational session data to make final conclusions.

## ACKNOWLEDGMENTS

The authors would like to thank Dr. *Phil Edwards* from the Institute of Space and Astronautical Science (Kanagawa, Japan) for an access to the calibration data of the VSOP - VLBA experiments. We also thank Dr. Sergey Likhachev (Astro Space Center) for his assistance during the data processing and very useful discussion.

This research was supported by the Russian Foundation of Basic Research through the *Grant No. 09-02-083563*